\documentclass[letterpaper]{article} 
\usepackage{aaai24}  
\usepackage{times}  
\usepackage{booktabs} 
\usepackage{caption}  
\usepackage{helvet}  
\usepackage{courier}  
\usepackage[hyphens]{url}  
\usepackage{graphicx} 
\usepackage{natbib}  
\usepackage{caption} 
\usepackage{algorithm}
\usepackage{algorithmic}
\usepackage{float}
\usepackage{lipsum,graphicx}
\usepackage{cuted}
\usepackage{subcaption}
\usepackage{amsmath}
\usepackage{newfloat}
\usepackage{listings}
\DeclareCaptionStyle{ruled}{labelfont=normalfont,labelsep=colon,strut=off} 
\floatstyle{ruled}
\newfloat{listing}{tb}{lst}{}
\floatname{listing}{Listing}
\title{Gene-associated Disease Discovery Powered by Large Language Models}
\author{
    Jiayu Chang$^1$\\
    Shiyu Wang$^2$ \\
    Chen Ling$^2$ \\
    Zhaohui Qin$^2$\\
    Liang Zhao$^2$
}
\affiliations{
Central South University$^1$, Emory University$^2$
}

\usepackage{bibentry}

\begin{document}

\maketitle

\begin{abstract}
The intricate relationship between genetic variation and human diseases has been a focal point of medical research, evidenced by the identification of risk genes regarding specific diseases. The advent of advanced genome sequencing techniques has significantly improved the efficiency and cost-effectiveness of detecting these genetic markers, playing a crucial role in disease diagnosis and forming the basis for clinical decision-making and early risk assessment. To overcome the limitations of existing databases that record disease-gene associations from existing literature, which often lack real-time updates, we propose a novel framework employing Large Language Models (LLMs) for the discovery of diseases associated with specific genes. This framework aims to automate the labor-intensive process of sifting through medical literature for evidence linking genetic variations to diseases, thereby enhancing the efficiency of disease identification. Our approach involves using LLMs to conduct literature searches, summarize relevant findings, and pinpoint diseases related to specific genes. This paper details the development and application of our LLM-powered framework, demonstrating its potential in streamlining the complex process of literature retrieval and summarization to identify diseases associated with specific genetic variations. 
\end{abstract}

\section{Introduction}
The correlation between genetic variation and human diseases is well-established. For instance, APOE-e4 is the first risk gene identified to remain strong risk on Alzheimer’s Disease (AD)~\citep{liu2013apolipoprotein}, HLA-Cw is one of genes that are most associated with psoriasis~\citep{huang2021hla}, and EGFR is known to be abnormal or mutated in nearly 50\% of lung cancers arising in those who have never smoked~\citep{bethune2010epidermal}. Therefore, identifying variations in disease-associated genes within the human genome is crucial for disease diagnosis~\citep{saunders2012rapid, zemojtel2014effective, martin2009gene}. The advances in genome sequencing techniques have made the detection of specific genetic markers or mutations more efficient~\citep{shih2023efficient, zhou2018whole} and cost-effective~\citep{christensen2015assessing, schwarze2018whole}. This can serve as crucial evidence for doctors in making clinical decisions, and provide valuable insights for early diagnosis and risk assessment, thereby enabling more informed and effective healthcare strategies. 
\begin{figure*}[t]
\centering
\includegraphics[width=0.85\textwidth]{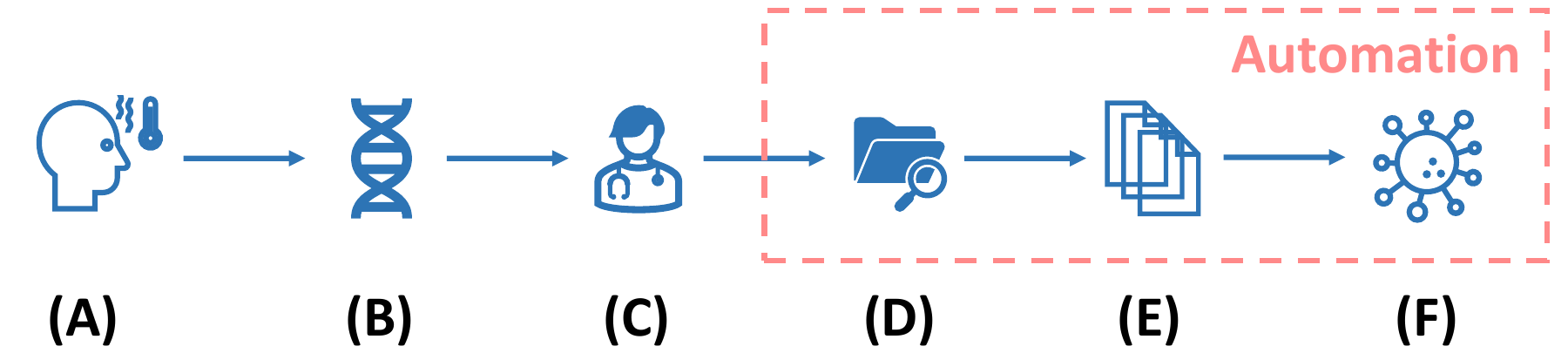}
\caption{Depiction of disease discovery process in clinical practice. It begins with the patient (A) visiting a clinic and undergoing genetic sequencing (B). The physician (C) then analyzes the sequencing results to pinpoint suspicious genetic variations. Subsequently, the physician searches databases or medical literature (D) for records pertinent to these specific genes (E). Finally, the potential disease related to these genes is identified. Our framework is designed to automate the labor-intensive steps from (D) to (F).}
\label{fig:clinic}
\end{figure*}
Early disease diagnosis using genetic factors typically requires population-level analysis. In statistical genomics, for instance, differential expression analysis is employed to identify genes with varying expression levels between diseased individuals and a healthy control group, using hypothesis testing. However, conducting such population-level studies can be prohibitively expensive and often impractical for physicians due to challenges in recruiting a sufficient number of participants. Therefore, given the genome sequencing of a potential patient, physicians can directly search for databases that contain information regarding disease-gene associations collected from literature or historical studies, such as DisGeNET~\citep{pinero2020disgenet}, Alzheimer’s Disease Neuroimaging Initiative (ADNI)~\citep{petersen2010alzheimer} and Diseases~\citep{grissa2022diseases}. However, these databases do not offer real-time updates on the current research progress of the community and can often be outdated. Alternatively, as shown in Figure~\ref{fig:clinic}, the physician usually searches for evidence in the medical literature that somehow are relevant to the genetic variations of interest, then analyzes the evidence related to each of the variations and identify the potential disease the patient may have.

The task of sifting through literature for evidence is exceedingly laborious, given the potential existence of thousands of papers concerning a specific gene. The researcher is tasked with the meticulous job of pinpointing those documents that specifically contain insights demonstrating the association of the gene with a particular disease. This process demands significant time and attention to detail, as it involves discerning the most relevant and informative studies from a vast sea of academic research. Hence, as indicated in Figure~\ref{fig:clinic}, the automation of the evidence retrieval process, encompassing tasks like literature search, summary generation, and disease recognition, could substantially enhance the efficiency with which physicians identify diseases related to specific genes. The recent advancements in Large Language Models (LLMs) have notably enhanced their ability to summarize scientific literature~\citep{yu2022evaluating, ghadimi2022hybrid}. This makes them well-suited for condensing literatures related to specific genes, thereby extracting vital information that can aid physicians in making informed clinical decisions. To accomplish this objective, we designed a framework powered by LLMs for discovering diseases associated with specific genes. This framework is capable of conducting literature search based on specified genes, summarizing the retrieved literatures, and identifying diseases related to the input genes. Utilizing this framework, the extensive and complex process of literature retrieval and summarization to identify potential disease from specific genes can be significantly streamlined and automated.

In this article, we begin by presenting an overview of related works, encompassing studies on disease-gene associations, databases that catalog these associations, and the use of LLMs in summarizing relevant literature. Following this, we will introduce our proposed framework, which utilizes LLMs for the discovery of diseases associated with specific genes. Subsequently, we will apply this framework to identify potential diseases linked to certain genes and delve into a discussion of the results obtained from these experiments.
\begin{figure*}[t]
    \centering
    \includegraphics[width=\textwidth]{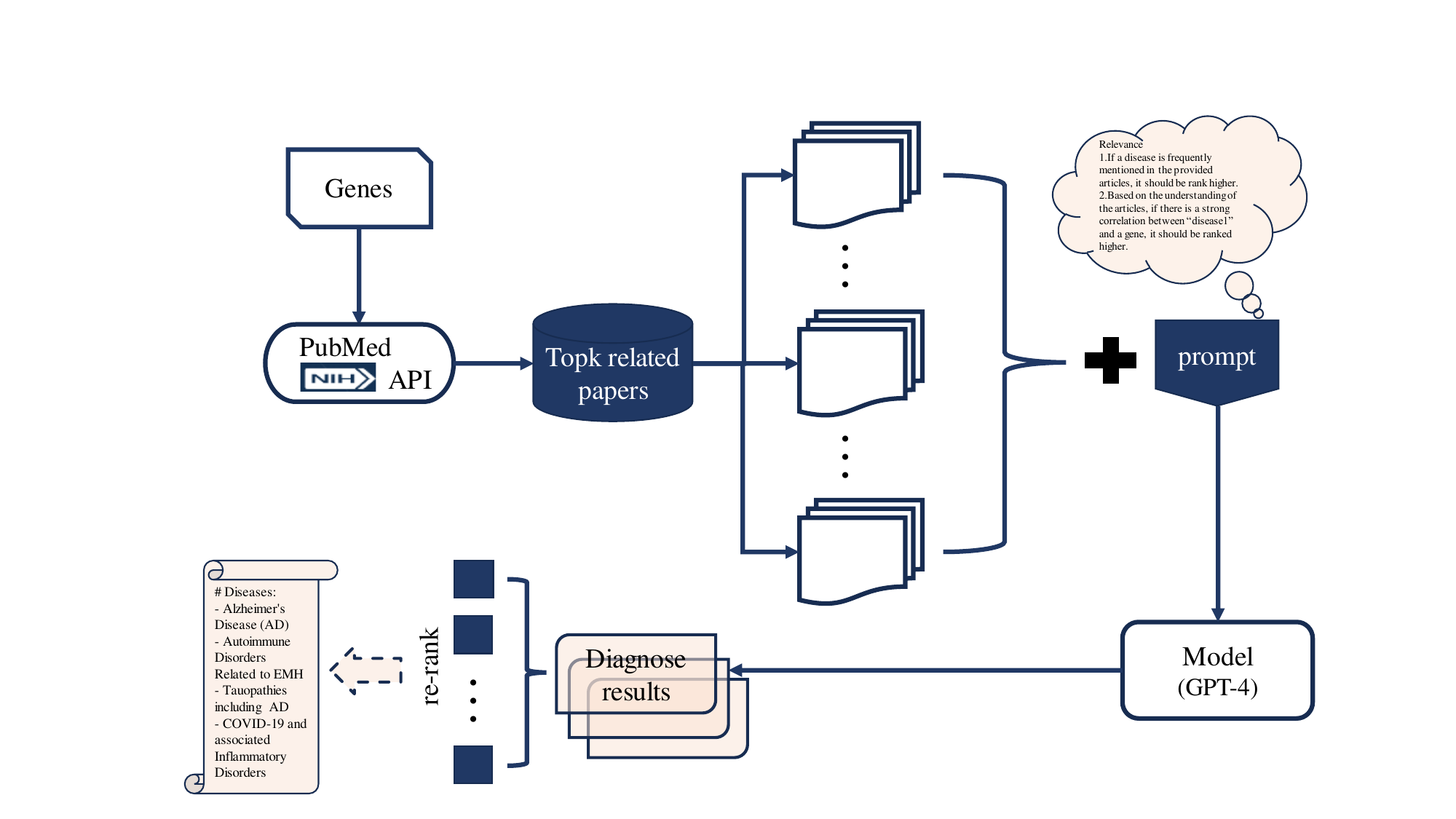}
    \caption{Framework of proposed method. The framework starts from specific genes suspicious to cause disease of the patient. Then the PubMed API is leveraged to search literatures regarding these genes by criteria such as relevance or time. Top K papers are then selected and queried based on crafted prompts by LLMs (e.g., GPT-4). During this phase, the content of the literature is analyzed by LLMs. Relevant diseases are identified and ranked through the in-context learning capabilities of Large Language Models LLMs. This process is iterated several times, with diseases being re-ranked based on the frequency of their occurrence in the outputs.}
    \label{fig:fig3}
\end{figure*}

\begin{figure}[t]
\centering
\includegraphics[width=0.5\textwidth]{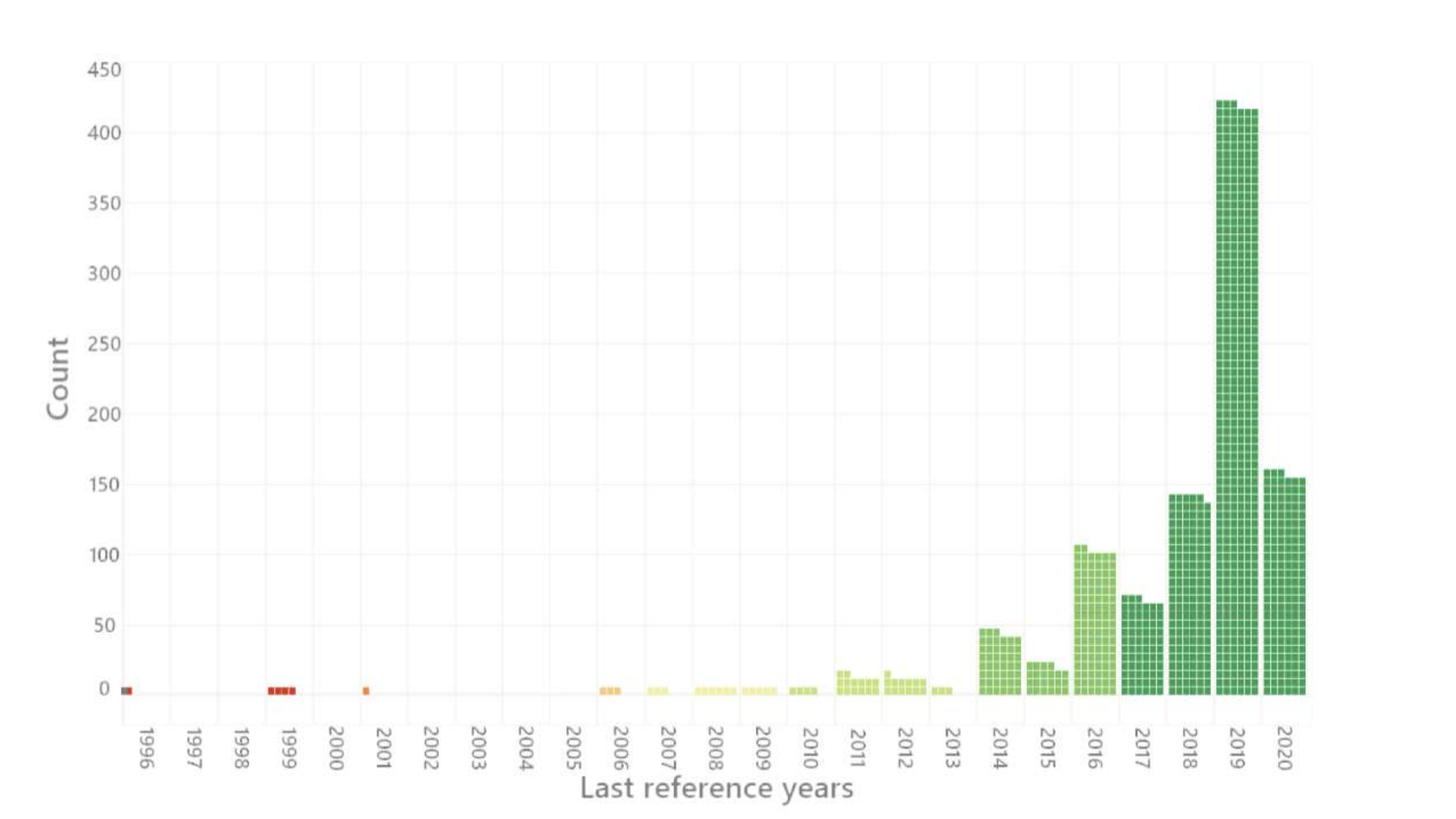}
\caption{Distribution of last-referenced years of selected genes in the dataset. The horizontal axis represents the most recent year a specific gene is referenced. The vertical axis represents frequencies.}
\label{fig:lastref}
\end{figure}
\section{Related works}
\label{sec:related}
\subsection{Disease-gene association studies}
High-throughput sequencing, particularly RNA-Seq, has emerged as the primary method for assessing expression levels~\citep{mortazavi2008mapping}. 
Following this, a range of methodologies for disease-gene association research have evolved, focusing on uncovering both diseases potentially connected to specific genetic variations and the underlying genetic processes involved in disease pathologies. Accurately identifying genes that are differentially expressed across specific conditions is crucial for understanding the variations in the occurrence of disease. For instance, baySeq~\citep{hardcastle2010bayseq} employs an empirical Bayes framework to identify patterns of differential expression in a collection of sequencing samples. It does this by assuming that the data follows a negative binomial distribution and by deriving a prior distribution that is empirically determined from the entire dataset. DESeq leverages negative binomial distribution, with variance and mean bound by local regression~\citep{anders2010differential}. edgeR utilizes a Poisson super dispersion model to address both technical and biological variations. It then employs a Bayesian empirical approach to moderate the level of overdispersion across transcripts~\citep{robinson2010edger}. After completing the differential expression analysis and fitting the model, the risk associated with the occurrence of a specific disease can be calculated for a given gene.

\subsection{Databases for disease-gene associations}
The ability to retrieve results from disease-gene association studies efficiently and conveniently is essential for their practical application in clinical settings. To facilitate this, several databases have been established to compile results from differential expression analyses. For example, DisGeNET is a discovery platform containing comprehensive publicly available collections of genes and variants associated to human diseases~\citep{pinero2020disgenet}, including 1,134,942 gene-disease associations between 21,671 genes and 30,170 diseases, disorders, traits, and clinical or abnormal human phenotypes. The latest update to these databases occurred in 2021, which could result in the omission of more recent studies. The Alzheimer’s Disease Neuroimaging Initiative (ADNI) database provides both cross-sectional and longitudinal characterization of clinical measures in individuals with mild AD and normal controls, alongside their sequencing data~\citep{petersen2010alzheimer}. This database also does not receive regular updates, which means that it may not include the most recent studies on AD. Diseases is a database that is updated weekly, encompassing disease-gene associations derived from text mining and data integration~\citep{grissa2022diseases}. Despite its weekly updates, this database is not capable of providing real-time retrieval of disease-gene associations and requires significant maintenance efforts. In contrast, our method achieves real-time retrieval, ensuring that physicians have access to the latest search results. This framework utilizes LLMs and does not require additional maintenance efforts.

\subsection{Literature summarization via LLMs}
The progress in text mining technology has enhanced the accuracy of summarizing specific topics into a concentrated version from multi-documents while keeping the main information~\citep{rahimi2017overview, abualigah2020text}. Databases used for retrieving disease-gene associations, such as Diseases~\citep{grissa2022diseases}, benefit from text mining-based methodologies and offer regular updates to their database content. Contrasting with traditional text mining methods that are limited by the length of texts and expressiveness, approaches based on LLMs are capable of processing long documents and offering more accurate summarizations. \cite{pilault2020extractive} demonstrated that transformer language models are surprisingly effective at summarizing long documents, outperforming typical seq2seq approaches. Moreover, in a specific study~\citep{goyal2022news}, participants demonstrated a strong preference for summaries generated by an Instruct-tuned 175B GPT-3 model, across two distinct styles using varied prompts. The study also revealed that GPT-3 summaries are of exceptionally high quality and are versatile enough to adapt to various summarization contexts. This insight has inspired our framework to harness the capabilities of pre-trained LLMs for multi-document summarization, enabling us to obtain high-quality insights about potential diseases associated with specific genes.

\section{Methods}
\label{sec:method}
As is shown in Figure \ref{fig:fig3}, our framework is mainly divided into four parts: gene-related literature retrieval, literature concatenation, literature summarization, and relevance ranking. In this section, we will introduce these parts in detail.

\subsection{Gene-Related Literature Retrieval}
First, we extracted gene information from the dataset. Then we utilized the PubMed API for literature retrieval. Because PubMed is a vast database covering life science and medicine, it provides a wealth of scientific and up-to-date resources. By invoking the PubMed API, we could retrieve literatures related to the genes under study from this database. Furthermore, a parameter K was set to determine the topK most relevant literatures we aimed to retrieve. By sorting the PubMed search based on relevance, we selected the topK literatures for further analysis.

\subsection{Literature Concatenation \& Summarization}
Initially, our endeavor involves the comprehensive preservation of the entirety of a given literature, encompassing heterogeneous unstructured data such as tables and images. Leveraging the vector storage tools within Langchain's vector space, we transform this voluminous content into vectors and subsequently concatenate the resultant documents. In order to forestall the potential breach of document length constraints, a segmentation process is employed, wherein the document is partitioned into discrete blocks. Subsequently, we employ Langchain for summarization.

However, this method exhibits sluggish performance when summarizing literature pertaining to individual gene datasets. The protracted nature of the data impedes the ability of the LLMs to discern salient points and comprehend the entirety of the article. Moreover, it introduces a propensity for erroneously associating unrelated diseases mentioned within the article as pertinent to the specific gene.

Upon careful scrutiny of pertinent literature, we observed a prevalent tendency wherein literatures concerning genes and diseases overtly delineate the principal research focus and content within the abstract section. Alternatively, they may explicitly denote the relevance relationship between the gene and disease, potentially substantiated through statistical metrics such as p-values. Consequently, for the specific task of literature retrieval from designated genes and subsequent summarization to discern potential diseases, we selectively retain only the abstract of each literature in local text format, concatenating them seamlessly. This strategic refinement markedly curtails the character count inputted to the LLMs, thereby mitigating redundancy and obviating the imperative to segment and warehouse literature in vectorized form. This approach optimizes the exploitation of the LLM's in-context learning prowess, thereby fortifying its grasp of crucial information within the realm of literature.

\subsection{Relevance Ranking}
Naturally, the greater the number of diseases retrieved as relevant to a gene, the more closely associated the gene is presumed to be with those diseases. This association is strengthened either by a higher frequency of appearance in the literature or by a notable emphasis on the gene-disease relationship within the text. Judgments regarding the correlation between genes and diseases are effectively drawn from both the frequency of occurrences and the model's attention to critical content. Both aspects hinge on the LLMs' in-context learning ability within the literatures.

To ensure a rigorous evaluation, we constrain the LLMs by configuring prompts (As shown in Figure \ref{fig:fig3}), limiting its information retrieval solely to the provided literature. This approach resembles a form of rejection sampling, preventing the model from relying on an internal knowledge base. Moreover, we impose restrictions on the ranking process based on the two aforementioned criteria.

To enhance the reliability of our results, we perform three queries to GPT for a given gene, adjusting the ranking based on the frequency of appearance in the results related to a specific disease. The final ranking is determined by sorting the diseases in descending order according to their respective frequencies across the three queries.

\begin{figure}[t]
    \centering
    \includegraphics[width=0.5\textwidth]{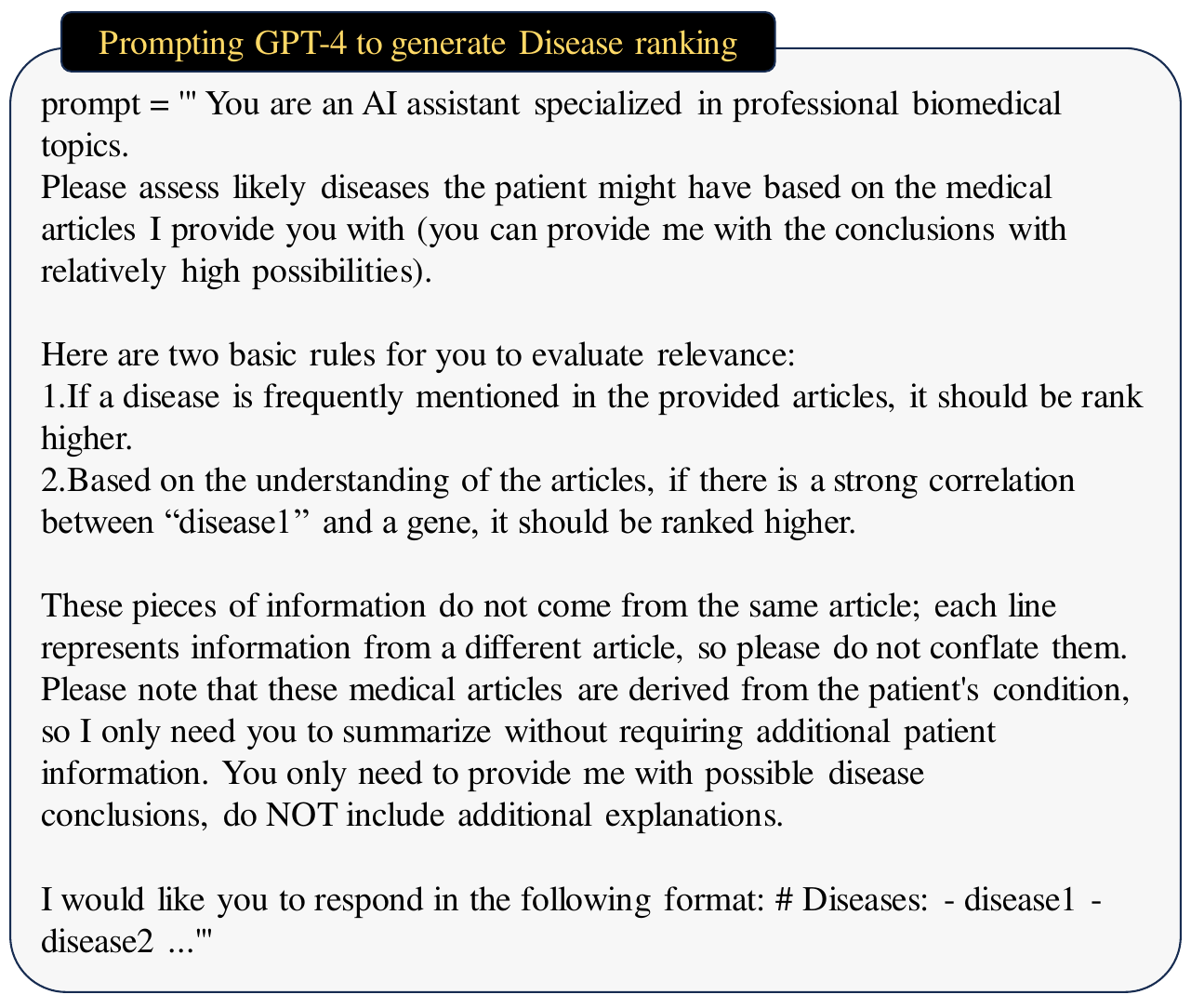}
    \caption{Prompt to instruct GPT-4 for disease ranking generation.}  
    \label{fig:fig3}
\end{figure}

\begin{table*}[t]
\centering
\captionsetup{justification=centering, singlelinecheck=false} 
\begin{tabular}{ccc}
\toprule
\textbf{Number (N)} & \textbf{10} & \textbf{20} \\
\midrule
\textbf{Percentage} & 24.59\% & 59.55\% \\
\bottomrule
\end{tabular}
\caption{Evaluating performance of the proposed framework by the percentage of the target disease that appears in retrieved results when N articles are used.}
\label{tab:1}
\end{table*}
\begin{table*}[t]
\centering
\captionsetup{justification=centering, singlelinecheck=false} 
\begin{tabular}{cccccc}
\toprule
\textbf{Number (N)} & \textbf{HR@2} & \textbf{HR@5} & \textbf{HR@10} & \textbf{HR@15} & \textbf{HR@20} \\
\midrule
\textbf{HR (N=10)} & 6.83\% & 16.39\% & 23.32\% & 24.39\% & 24.59\% \\
\textbf{HR (N=20)} & 13.76\% & 28.07\% & 53.90\% & 58.38\% & 59.55\% \\
\bottomrule
\end{tabular}
\caption{Evaluating performance of the proposed framework by the hit ratio (HR) regarding the target disease that appears in top N ranked retrieved results.}
\label{tab:2}
\end{table*}
\section{Experiments}
\label{sec:exp}
In this section, we will introduce the evaluation of the proposed framework. We will firstly introduce the dataset borrowed in the experiments, followed by the evaluation metrics. Then we demonstrates the experimental results with a comprehensive discussion. In the end, we will showcase a case study by illustrating input genes, ranking results of retrieved diseases and highlighted evidence in the literature.
\subsection{Dataset}
To evaluate our framework, we select 1,025 genes that are know to be associated with AD based on the existing record from  DisGeNET~\citep{pinero2020disgenet}. The latest reference of those genes ranges from 1996 to 2020, based on the record of DisGeNET. Ideally, we use these genes as inputs for our framework, anticipating that the output will include AD among its results. For genes that were mentioned very early on (e.g., in 1996), they might not be the current focus of the research community. As shown in Figure~\ref{fig:lastref} that illustrates the distribution of last-referenced years of these genes, only few genes were referenced before 2010, but consequently our framework may not provide detailed summarizations for these genes, owing to the limited number of documents that LLMs can process.
\begin{figure}[t]
\centering
\subcaptionbox{Frequency of rank of Alzheimer's Disease in the output (K=10)}{\includegraphics[width=0.45\textwidth]{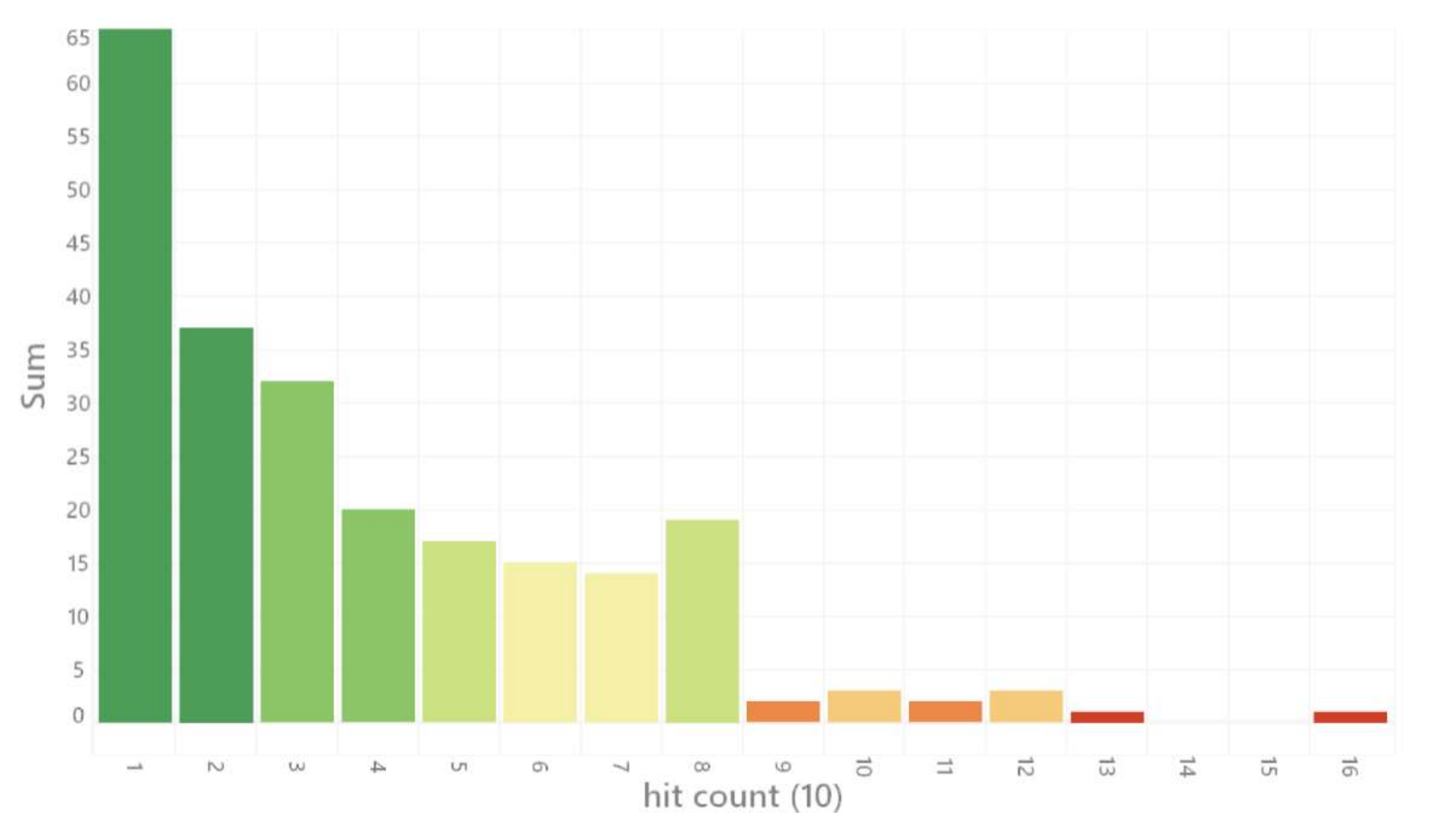}}
\subcaptionbox{Frequency of rank of Alzheimer's Disease in the output (K=20)}{
    \includegraphics[width=0.45\textwidth]{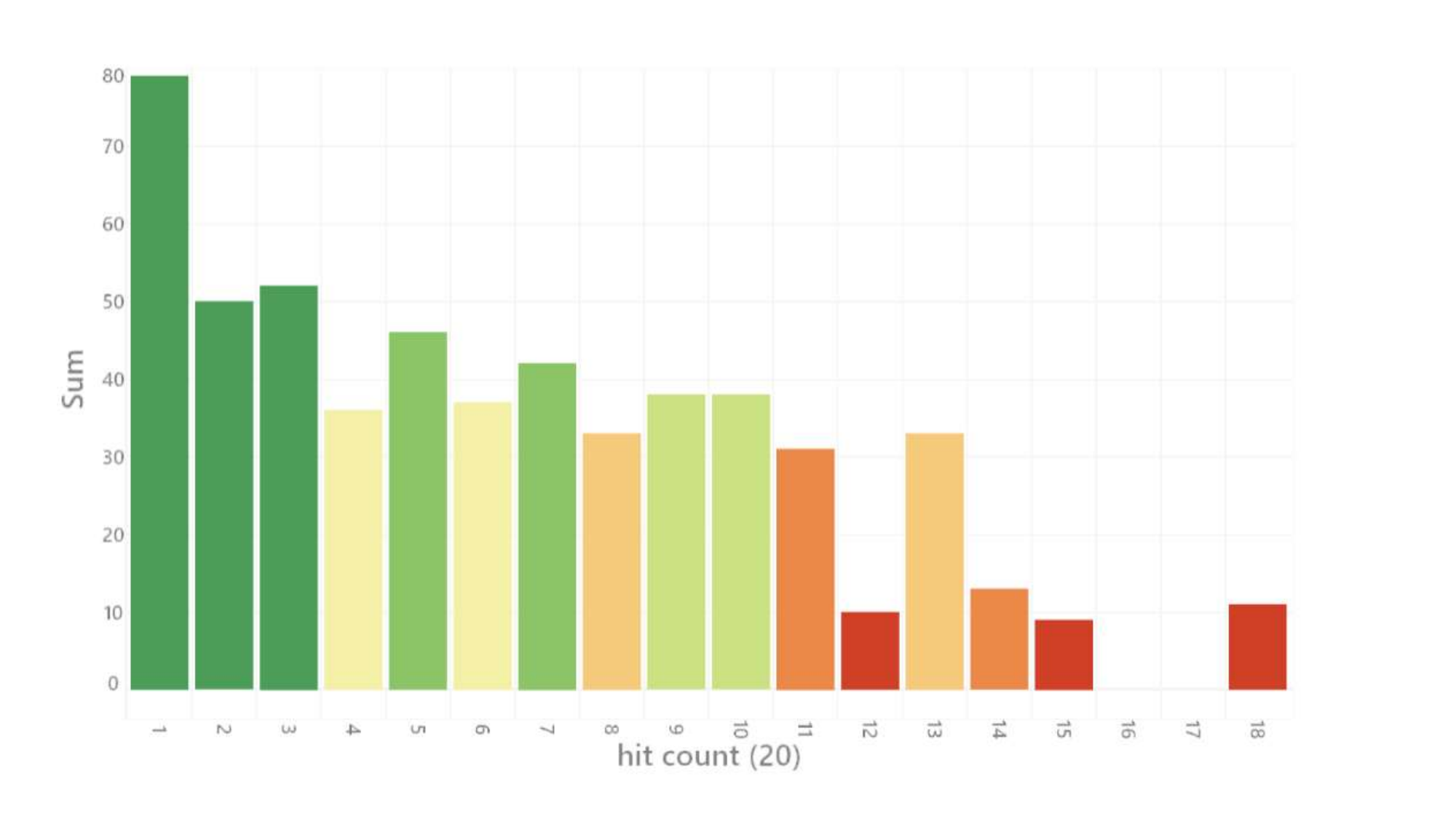}
}
\caption{Distribution of rank of Alzheimer's Disease in the output}
\label{fig:4-9}
\end{figure}
\begin{figure*}[t]
    \centering    \includegraphics[width=\textwidth]{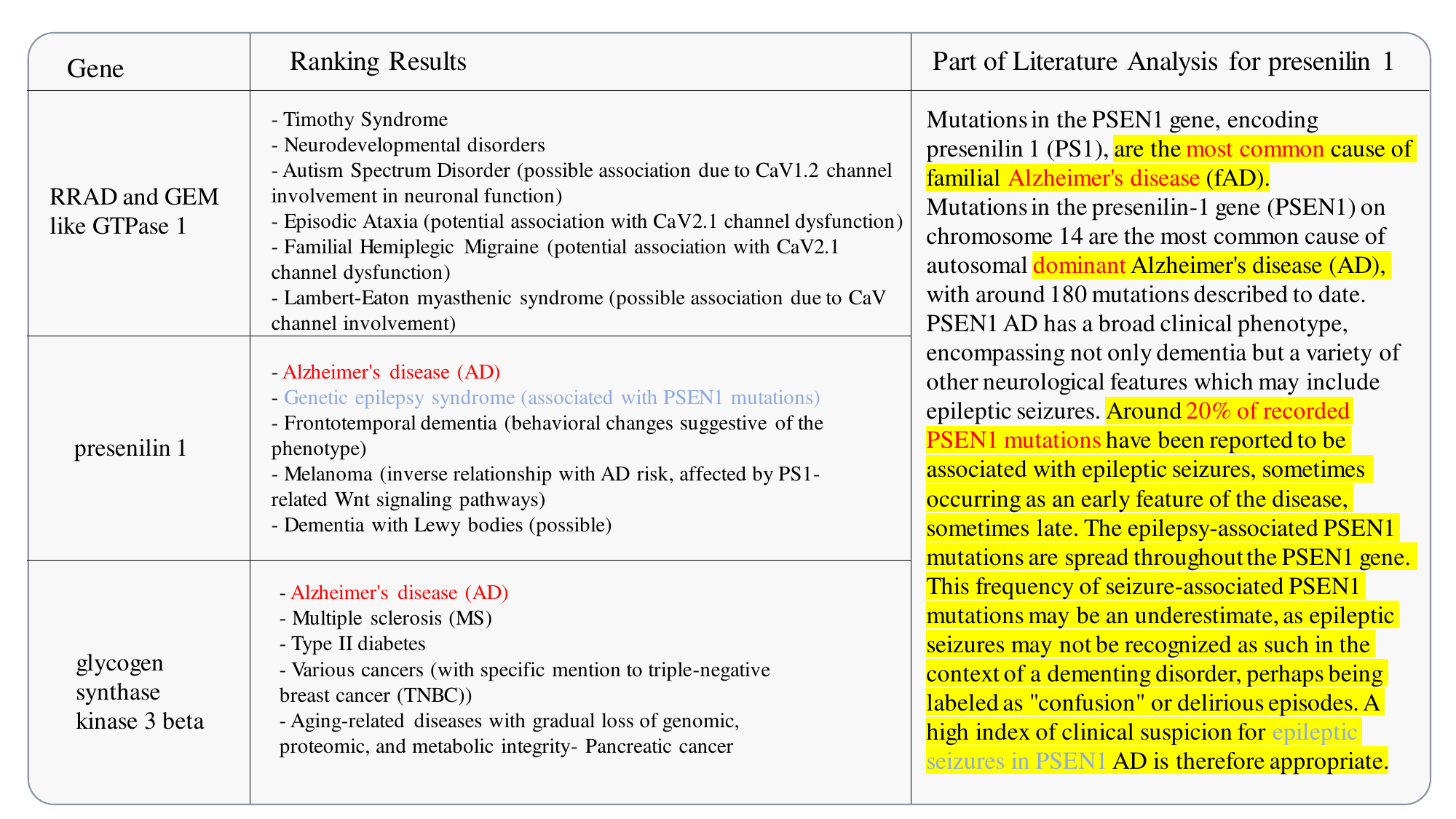}
    \caption{Analysis of samples from our framework given specific genes}
    \label{fig:fig5}
\end{figure*}
\subsection{Evaluation Metrics}
We used the Hit Ratio (HR), a commonly used metric for measuring recall in top-k recommendations.
As shown in Equation (1):
\begin{equation}
    \text{HR} = \frac{\text{NumberOfHits@K}}{\text{GT}}
    \label{1}
\end{equation}

Where: NumberOfHits$@$K is the number of successfully hit instances of Alzheimer's Disease.
GT is the sum of the counts of gene-associated diseases belonging to the test set in the top-K lists for each gene.

\subsection{Results}
Table \ref{tab:1} indicates that with the increase of N, the frequency of Alzheimer's disease appearing in the ranking list is on the rise. This suggests that by further expanding the retrieval scope, for instance, with N=30,40, there is still room for an increase in the probability of Alzheimer's disease occurrence. Alternatively, when transitioning our method to a larger and more comprehensive database base, there may be an even more substantial improvement in performance. This underscores the effectiveness of our approach.

Table \ref{tab:2} shows the HR at the top 2, 5, 10, 15, and 20 for the specified gene. Through the tuning of the hyperparameter N in literature retrieval, it is observed that with an increase in the number of accessible literature references and the intensification of knowledge density, the HR corresponding to the target disease, Alzheimer's disease, gradually rises. This aligns well with the first criterion for determining relevance: an increased frequency of papers related to the gene-disease association indicates a higher level of attention from medical researchers in the specialized field. Consequently, the greater the frequency of occurrences, the stronger the perceived relevance between the disease and the gene.

It is noteworthy that when N is set to 10, there are variations in the HR at the top 10, 15, and 20. This phenomenon arises because, even with only 10 literature references available for LLMs reference, certain articles may mention correlations between the gene and multiple diseases in their abstracts. However, this study specifically explores a singular disease. This results suggests that LLMs does indeed demonstrate a certain level of comprehension of the content in the articles. Through the effective utilization of its in-context learning capability, LLMs is able to understand the relationships between the mentioned gene and disease in the literature, rather than merely processing medical terminology without context.

\subsection{Case Study}

Figure \ref{fig:fig5} demonstrates a case study related to genes, where we present two genes highly associated with Alzheimer's disease, and the corresponding literature sections for the top two diseases regarding one of these genes. As shown in Figure~\ref{fig:fig5}, LLMs is able to detect and summarize disease relevant to the input genes from articles retrieved by PubMed API. For instance, for gene presenilin 1, GPT-4 leverages information in the article (highlighted in yellow) such as "Mutations in the PSEN1 gene, encoding presenilin 1 (PS1), are the most common cause of familial Alzheimer's disease (fAD)." and "Around 20\% of recorded PSEN1 mutations have been reported to be associated with epileptic seizures..." to determine the list of disease as the output and rank AD on top of genetic epilepsy syndrome. The example demonstrates that LLMs (e.g., GPT-4) can effectively comprehend medical-related content and capture terms such "most common," "20\%," etc., to assess the strength of correlation between genes and diseases.

\section{Conclusion}
\label{sec:conclusion}
In conclusion, our framework automates the labor-intensive process of discovering diseases associated with specific genes. By conducting literature retrieval and summarizing relevant findings, we enhances the efficiency of disease identification, offering a valuable tool for medical research. We conducted experiments on the selected DisGeNET dataset, and the results clearly indicate that our framework is indeed capable of effectively ranking diseases based on the strength of their correlation with genes.

In future work, we aim to develop and implement advanced retrieval techniques that efficiently amalgamate and process data from various sources, such as text and images. This multi-modal approach will enhance the ability of our system to access and interpret a wide range of information. Moreover, we plan to design and test nuanced ranking algorithms. These algorithms will be tailored to effectively cater to the unique requirements of different applications, ensuring that our system can provide relevant and accurate results across a variety of use cases.

\bibliography{aaai24}

\end{document}